\documentclass[aps,prd,reprint,showpage,showpacs,nobibnotes,superscriptaddress,twocolumn,amssymb,amsmath,nofootinbib,floatfix]{revtex4-1}
\usepackage{graphicx}
\usepackage[usenames,dvipsnames]{xcolor}
\usepackage{epsfig}
\usepackage{verbatim}
\usepackage{amsthm}
\usepackage{amsfonts}
\usepackage{amscd}
\usepackage{bm}
\usepackage[caption=false]{subfig}
\definecolor{linkcolor}{RGB}{55,57,154}
\usepackage[pdfstartview={XYZ},colorlinks=true,allcolors=linkcolor]{hyperref} 
\begin{document}
\title{Type I + III seesaw mechanism and CP violation for leptogenesis}
\date{\today}
\author{Edison T. Franco}
\email{edisonfranco@uft.edu.br}
\affiliation{Universidade Federal do Tocantins, Campus Universit\'ario de Aragua\'\i na \\
Av. Paraguai, 77814-970 Aragua\'\i na, TO, Brazil}

\begin{abstract}
 A seesaw mechanism is presented in the neutrino sector and a new phase of CP violation ($\alpha$) emerges in the interplay between the type-I and type-III seesaw schemes. This phase is inside the mixing term, and thus it cannot be rotated away in the Yukawa Lagrangian and, therefore, the heavy symmetry states cannot be in a diagonal weak basis in the broken phase. Some particular descriptions are analyzed suggesting that if the usual Yukawa couplings are suppressed, leptogenesis still occurs due to a new interacting vertex with fermion triplet $T$, fermion singlets $N$, and an ad-hoc scalar triplet, $\Sigma$, which now is included to mediate the interactions. The evaluated CP violation is enough to generate the observed matter-antimatter asymmetry even in the minimal $1N+1T$ case (independently of $\alpha$) or in the $2N+1T$ approach (controlled by $\alpha$). The latter introduces more CP contributions to leptogenesis due to new diagrams which are now possible even with the suppressed imaginary part of the standard Yukawa couplings and can induce the observed baryon-to-photon ratio.
\end{abstract}

\maketitle

\section{Introduction}

\label{sec:intro} Minimal extensions of the standard model (SM) usually add heavy singlet fermions
(commonly called as right-handed neutrinos) to the SM lepton sector, $N_{iR}\sim \left( 1,0\right) $. They have
very large Majorana masses and are linked to the light neutrinos realizing
the type-I seesaw mechanism \cite{seesaw:I} and also producing the lepton
asymmetry,
which is transferred to the visible baryon sector, generating the baryon asymmetry of universe (BAU)~\cite{Sakharov:1967dj}. It occurs due to sphaleron processes~%
\cite{sphaleron} in the so-called leptogenesis mechanism~\cite{Fukugita:1986hr,Davidson:2008bu,Branco:2011zb}. The existence of similar
mechanisms of type-II~\cite{seesaw:II} and type-III~\cite{Foot:1988aq} adds
to the standard model scalar triplets, $\Sigma \sim \left( 3,+2\right) ,$ or
right-handed fermion triplets, $T_{R}\sim \left( 3,0\right) ,$ respectively. Nevertheless, the hypercharge forbids their general interaction. These usual seesaw mechanisms
describe a Majorana mass term for the neutrinos as $\mathcal{L^{\textrm{mass}}}=\frac{1}{2}n_{L}^{T}\mathcal{C}\mathcal{M}^*n_{L}+H.c.$, where the neutrino mass matrix is constructed on the symmetry basis $n_L=\left(
\begin{array}{ccc}
\nu _{L} & N_{R}^{c} & T_{R}^{0 c}%
\end{array}%
\right) ^{T}$ and the fields $\nu _{L}
=\left( \nu _{1L},\nu _{2L},\nu _{3L}\right) ^{T}$ are the
left-handed neutrino components, $N_{R}^{c}\equiv (N_R)^{c}=\left( N_{1R}^{c},N_{2R}^{c},N_{3R}^{c}\right) $ are the
heavy right-handed singlet components and $T_{R}^{0 c}\equiv (T_{R}^{0})^{c}=\left(
T_{1R}^{c},T_{2R}^{c},T_{3R}^{c}\right) $ are the neutral fermion triplet components in the most general three fermion family model.
The full diagonalization of the general $\mathcal{M}_{9\times 9}$ allows one to obtain the
physical neutrino eigenstates and generally carries enough phases to CP at low
energies even in minimal setups~\cite{Branco:2002xf}. Minimal versions of these extensions with only two heavy neutral fermions (general $\mathcal{M}$ with only $5\times5$ entries) have been extensively studied in the literature~\cite{Frampton:2002qc} and are widely applied to the leptogenesis mechanism.

Since in the standard scheme there is nothing to forbid the use of a weak basis (WB) where
the masses of heavy neutrinos ($\mathbf{M}_{\mathrm{N}}$) and heavy triplets ($\mathbf{M}_{\mathrm{T}}$) are real and diagonal, one can choose this WB to easily diagonalize $\mathcal{M}$. Once the components of $\mathbf{M}_{\mathrm{N}}$
and/or $\mathbf{M}_{\mathrm{T}} $ are much heavier than the Dirac off-diagonal components ($\mathbf{m}_D$ and $\mathbf{m}_D^{\prime}$) one can
make a block diagonalization and, in very good approximation \cite%
{Branco:2001pq}, the light neutrino sector has the non diagonal masses given by $\mathbf{m}_{\nu }\simeq \mathbf{m}{_{_{\textrm{II}}}}-\mathbf{mM}^{-1}\mathbf{m}^{T}$,
where $\mathbf{m}=(
\mathbf{m}_{D},\mathbf{m}_{D}^{\prime })$ and  $\mathbf{M}=\textrm{diag}(
\mathbf{M}_{\mathrm{N}} ,\mathbf{M}_{\mathrm{T}} )$. The full diagonalization is easily made since $D=\mathbf{M}$ in this WB
while the light neutrinos have the masses diagonalized by $%
d=\mathbf{K}^{\dagger }\mathbf{m}_{\nu }\mathbf{K}^{\ast }$,
where $\mathbf{K}$ is the Pontecorvo-Maki-Nakagawa-Sakata (PMNS) mixing matrix.
Even if we include more than one
singlet and/or triplet, the structure of $\mathbf{M}$ will still be the same,
with $\mathbf{M}_{\mathrm{N}} $ and/or $\mathbf{M}_{\mathrm{T}} $ being the related diagonal submatrices.
The off-diagonal zeros persist unless new content is added to the model.

Seesaw models introduce minimal modifications to SM content by some new particles. Here we need to go beyond and insert an interaction between these new species. In the minimal version at low energies, for
example, the interaction between singlets and the
neutral scalar component can generate nonzero matrix elements if this
interaction receives a mass term from a vacuum expectation value (VEV) of a new \textit{ad hoc} (nonusual) scalar
triplet, $\Sigma \sim \left( \mathbf{3},0\right) $, which can interact with $N_{R}$
and $T_{R}$ and, therefore, produces a new seesaw mechanism by the
introduction of off-diagonal terms into $\mathcal{M}$ (specifically into the submatrix $%
\mathbf{M}$). This hybrid interaction of the types I and III seesaw mechanisms emerges naturally from the unification when the neutrino mass problem is solved by the inclusion of an adjoint fermion representation in the minimal $SU(5)$~\cite{Bajc:2006ia}. The adjoint representation has a branch to the SM group, $SU(3)\otimes SU(2) \otimes U(1)$, given by $\mathbf{24}=(1,1)_0+(1,3)_0+(3,2)_{-5/6}+(\bar{3},2)_{5/6}+(8,1)_0$. Hence, an interaction term as $\mathcal{L}^{\mathrm{int}}\ni N_R Tr(T_R \Sigma)$ can be naturally achieved from an interaction between two fermions and a scalar adjoint representation, like $\mathrm{Tr}(\mathbf{24}_F \mathbf{24}_F \mathbf{24}_S)\ni((1,1)_{0F}(1,3)_{0F}(1,3)_{0S}))$~\cite{Bajc:2006ia,Kannike:2011fx}. This content is customarily necessary to save the unification in a variety of models based on $SU(5)$~\cite{su5}.

Besides the similar aspects of type I and type III seesaw
mechanisms, the introduction of this peculiar interaction generates new
interferences between the tree and loop levels, which contribute to raise the CP violation in the lepton sector.
Yet, the interplay of the different seesaw mechanisms may be important for the neutrino mass generation and should explain the small deviations from the tribimaximal mixing form of the PMNS matrix~\cite{Sierra:2013ypa}.

Such interactions can induce CP violation in both $N$ and $T$ decays and should be driven by a new CP source. Moreover, previous studies have shown that the second heavy eigenstate may be important for leptogenesis in the thermal scenario~\cite{DiBari:2005st}. The asymmetry can be associated to $N$ (or $N_1$), which is heavier than $T$, however, playing the key role for CP violation to leptogenesis. From this point of view, all CP violation generated by the two lightest eigenstates among all heavy neutral fermions should be taken into account. On the other hand, the scalar triplet $\Sigma$ can naturally get a vanishingly small VEVV since it may be inside of the $24_S$ in the unification in $SU(5)$~\cite{Bajc:2006ia}. Thus, the study of CP in the decay of the new fermions to $\Sigma$ may be important if an interaction between a fermion singlet, fermion triplet, and scalar triplet is included. It provides a new relevant CP origin and complements the usual lepton asymmetries calculated in the standard seesaw mechanisms in leptogenesis~\cite{Davidson:2008bu}.

To study the consequences of this kind of models we will focus on the
minimal examples with only $1N_R$ and $1T_R$ (1N1T) and with $2N_R$ and $1T_R$ (2N1T)
. In this vein, the paper is organized as follows. In Sec. \ref%
{sec:minimal_typeIV} we introduce the minimal $1N1T$ type I+III seesaw model and some
aspects of the diagonalization mechanism for the Majorana sector. In Sec.~\ref{sec:CP1N1T} we show the CP violation generated in $T$ and  $N$ decays for the 1N1T setup before $\Sigma$ gets a VEV (diagonal $\mathbf{M}$ mass matrix). In Sec. \ref{sec:CP2N1T} the CP
violation driven by $T$ and the $N_1$ decays is discussed in the 2N1T case (also
in the symmetric phase) and some results are derived to generate the baryon-to-photon ratio in this case.
Finally, our
conclusions and outlook are given in Sec. \ref{sec:conclusion}.

\section{Minimal Type I+III Seesaw Model}
\label{sec:minimal_typeIV}
For the following discussion let us concentrate on
the case where only one fermion singlet and one fermion triplet are added to SM. In this footing, the most general lepton sector invariant Lagrangian is given by%
\begin{eqnarray}
-\mathcal{L} &=&y_{i}^{\nu }\bar{\ell _{Li}}N_{R}\tilde{H}+\sqrt{2}%
y_{i}^{t}\bar{\ell _{Li}}T_{R}\tilde{H}+y^{_\Sigma }N_{R}^{T}\mathcal{C}%
\mathrm{Tr}\left( \Sigma T_{R}\right) \notag \\ && +\frac{1}{2}{M}_{\mathrm{N}} N_{R}^{T}\mathcal{C}%
N_{R} +\frac{1}{2}M_{\mathrm{T}} \mathrm{Tr}\left( T_{R}^{T}\mathcal{C}T_{R}\right) +\frac{%
1}{2}M_{\Sigma }^{2}\left\vert \Sigma \right\vert ^{2} \notag \\ &&
+V(H,\Sigma)
+\mathrm{H.c.},
\label{eq:LagMin}
\end{eqnarray}%
where the fermion and scalar triplets are, respectively, defined as
\begin{equation}
T_{R}=\frac{1}{\sqrt{2}}\left(
\begin{array}{cc}
T_{R}^{0} & \sqrt{2}T_{R}^{+} \\
\sqrt{2}T_{R}^{-} & -T_{R}^{0}%
\end{array}%
\right) \sim \left( \mathbf{3},0\right) ,
\end{equation}%
and
\begin{equation}
\Sigma =\frac{1}{\sqrt{2}}\left(
\begin{array}{cc}
\Sigma ^{0} & \sqrt{2}\Sigma ^{+} \\
\sqrt{2}\Sigma ^{-} & -\Sigma ^{0}%
\end{array}%
\right) \sim \left( \mathbf{3},0\right),
\end{equation}%
where the scalar potential $V(H,\Sigma)$ encodes all purely scalar interactions. The other lepton fields are the same as in the SM with heavy Majorana
neutrinos [$\ell \sim \left( \mathbf{2},-\frac{1}{2}\right) $, $N_{R}\sim
\left( \mathbf{1},0\right) $], where $H\sim \left( \mathbf{2}%
,\frac{1}{2}\right)$ is the scalar doublet and $\tilde{H}=i\sigma _{2}H^{\ast }$. The Lagrangian
in Eq.$( \ref{eq:LagMin})$ does not conserve lepton number in either $N_{R}$ or $T_{R}$ decays. After all spontaneous symmetry breakings, $H$ gets a VEV $v$ and $\Sigma$ gets $v_{_\Sigma}$. The
neutrino mass term is given by  $\mathcal{L^{\textrm{mass}}}=\frac{1}{2}n_{L}^{T}\mathcal{C}\mathcal{M}^*n_{L}+H.c.$, where the
general neutrino mass matrix, $\mathcal{M}$, is fulfilled with all seesaw
mechanisms and can be written as follows
\footnote{Notice that this is similar to the structure obtained in the inverse seesaw mechanism~\cite{Mohapatra:1986bd}.}:
\begin{equation}
\mathcal{M}=\left(
\begin{array}{ccc}
\mathbf{m}_{_{\textrm{II}}} & \mathbf{m}_{D} & \mathbf{m}_{D}^{\prime } \\
\mathbf{m}_{D}^{T} & {M}_{\mathrm{N}}  & {M}_{\textrm{Hy}} \\
\mathbf{m}_{D}^{\prime T} & {M}_{\textrm{Hy}}^{T} & {M}_{\mathrm{T}} %
\end{array}%
\right) .  \label{Miv}
\end{equation}%
In the present case, the entries of  the above $5\times 5$ matrix are submatrices given
by the following connection with the Lagrangian in Eq.~(\ref{eq:LagMin}):
\begin{equation}
\mathbf{m}_{_{\textrm{II}}}=\mathbf{0}_{3\times3}, \ \mathbf{m}_{\textrm{D}i}=%
\frac{vy_{i}^{\nu }}{\sqrt{2}}, \ \mathbf{m}_{\textrm{D}i}^{\prime }=%
\frac{vy_{i}^{t}}{\sqrt{2}}, \ {M}_{\textrm{Hy}}=\frac{v_{_\Sigma }y^{_\Sigma }}{%
\sqrt{2}},
\end{equation}%
where ${M}_{\textrm{Hy}}$ is a scalar complex mass term in the
minimal $1N1T$ model. Consequently, the heavy particles are not in the mass
eigenstates and we have to diagonalize $\mathcal{M}$ by an orthogonal
transformation. Let $\mathbf{V}$ be the matrix which does that,
\begin{equation}
\mathcal{D}\equiv\mathbf{V}^{T}\mathcal{M}^{\ast }\mathbf{V}=\mathrm{diag}\left(
m_{1},m_{2},m_{3},M_{1},M_{2}\right),
\end{equation}%
where the lowercase letters corresponds to light neutrino masses and
uppercase letters denote the two effective heavy neutrino masses. This type of minimal
setup usually induces a light neutrino to be massless~\cite{Frampton:2002qc}.

The diagonalization of $%
\mathcal{M}$ in Eq.~(\ref{Miv}) is similar to the one described in Ref.~\cite{Branco:2001pq}. Let us
rewrite $\mathcal{M}$ disconnecting the heavy and light sectors,
\begin{equation}
\mathcal{M}=\left(
\begin{array}{cc}
0 & \mathbf{m} \\
\mathbf{m}^{T} & \mathbf{M}%
\end{array}%
\right) ,
\end{equation}%
where $\mathbf{m}=(\mathbf{m}_{\textrm{D}},\mathbf{m}_{\textrm{D}}^{\prime })$ and $\mathbf{M}$ is a complex submatrix given by
\begin{equation}
\mathbf{M}=\left(
\begin{array}{cc}
{M}_{\mathrm{N}}  & {M}_{\textrm{Hy}} \\
{M}_{\textrm{Hy}} & {M}_{\mathrm{T}} %
\end{array}%
\right),
\end{equation}%
with real elements in the diagonal.
Assuming the form of $\mathbf{V}$ as in the type-I
seesaw mechanism~\cite{Branco:2001pq},
\begin{equation}
\mathbf{V}=\left(
\begin{array}{cc}
\mathbf{K} & \mathbf{R} \\
\mathbf{S} & \mathbf{T}%
\end{array}%
\right),
\end{equation}%
we find,
\begin{eqnarray}
\mathbf{d} &\simeq &-\mathbf{K}^{T}\mathbf{mM}^{-1}\mathbf{m}^{T}\mathbf{K},  \label{dn} \\
\mathbf{D} &\simeq &\mathbf{T}^{\dagger }\mathbf{M}\mathbf{T}^{\ast },  \label{DN}
\end{eqnarray}%
where we should not generally use $\mathbf{T}\approx \mathbf{1}$ since in this case the
nondiagonal entries can contribute in an unusual way depending on the strength of the coupling $y^{_\Sigma}$. On the other hand, it
is easy to show that $\mathbf{S}\sim \mathbf{M}^{-1}~\mathbf{m}$, $\mathbf{R}\sim \mathbf{%
m~M}^{-1}$, and (to a good approximation) $\mathbf{V}$ can be written in
terms of only two matrices, namely, $\mathbf{K}$ and $\mathbf{T}$, as follows:
\begin{equation}
\label{eq:vsig}
\mathbf{V}=\left(
\begin{array}{cc}
\mathbf{K} & \mathbf{mM}^{-1}\mathbf{T} \\
-\left( \mathbf{M}^{-1}\right) ^{\dagger }\mathbf{m}^{\dagger }\mathbf{K} & \mathbf{T}%
\end{array}%
\right).
\end{equation}
While $\mathbf{K}$ is linked to the PMNS matrix and diagonalizes the light neutrino mass matrix, $\mathbf{T}$ is the matrix which does the similar process in the heavy sector. The study of the connection of leptogenesis with light energy parameters has been extensively explored in the literature~\cite{Branco:2006ce} and we will not focus on it.
%

Let us assume a predictive
point of view in which we have $\left\vert {M}_{\mathrm{N}} \right\vert $,$\left\vert
{M}_{\mathrm{T}} \right\vert \gg \left\vert {M}_{{\mathrm{Hy}}}\right\vert $. Generically speaking, the
entries $M_{\mathrm{N}} $ and ${M}_{\mathrm{T}} $ can be complex constants; then, we can rotate
the lepton fields to a WB where these quantities are real and absorb all Majorana phases into $y^{_\Sigma
} $. Therefore, we will start considering $M_{\mathrm{N}} $ and ${M}_{\mathrm{T}} $ just as real
elements. Furthermore, these
phases cannot be rotated away by a redefinition of both Majorana fields, allowing for a complex $\mathbf{M}$. The matrix  $\mathbf{M}$ is correctly diagonalized if we define the Hermitian operator, $\mathbf{H}=\mathbf{MM%
}^{\dagger }$, through the unitary matrix $\mathbf{T}$ as $%
\mathbf{D}^{2}=\mathbf{T}^{\dagger }\mathbf{H}\mathbf{T}$. The leptonic CP phases are now in $%
\mathbf{m}$ and also in $\mathbf{M}$, as can be seen from Eqs.~(\ref{dn}) and~(\ref{DN}). The
four parameters in $\mathbf{M}$ are now manifested as two heavy
masses, $M_{1}$ and $M_{2}$, and two phases in the matrix $\mathbf{T}$, for which we choose the following
form:
\begin{equation}
\mathbf{T}=\left(
\begin{array}{cc}
\cos \theta & \mathrm{e}^{-i\phi }\sin \theta \\
-\mathrm{e}^{i\phi }\sin \theta & \cos \theta%
\end{array}%
\right) .
\label{Trans}
\end{equation}%
If we introduce the Majorana phase $\alpha $ by doing an explicit
phase extraction,
\begin{equation}
\label{eq:m4}
{M}_{{\mathrm{Hy}}}\equiv m_{\textrm{hy}}e^{i\alpha }=\frac{v_{\Sigma }}{\sqrt{2}}\left\vert y^{_\Sigma
}\right\vert e^{i\alpha },
\end{equation}
the general eigenvalues of $\mathbf{H}$ are given by
\begin{equation}
h_{\pm }=\frac{1}{2}\left( 2m_{\textrm{hy}}^{2}+M_{\mathrm{N}} ^{2}+M_{\mathrm{T}} ^{2}\pm r\right),
\end{equation}
where
\begin{equation}
r= \sqrt{\left(M_{\mathrm{N}} ^{2}-M_{\mathrm{T}} ^{2}\right) ^{2}+4m_{\textrm{hy}}^{2}
\left(M_{\mathrm{N}} ^{2}+M_{\mathrm{T}} ^{2}+2M_{\mathrm{N}} M_{\mathrm{T}} \mathrm{\cos }2\alpha\right)
}.
\end{equation}
Thus, the effective physical masses are given by $M_{1}=%
\sqrt{h_{-}}$ and $M_{2}=\sqrt{h_{+}}$. If we assume that $m_{\textrm{hy}}$ is suppressed compared to $M_{\mathrm{N}}$ and $M_{\mathrm{T}}$, then one gets $M_{1}\simeq M_{\mathrm{T}} $ and $M_{2}\simeq M_{\mathrm{N}}$ (exactly equal in the decoupled limit $m_\textrm{hy}\rightarrow 0$). More generically, it can
be expanded for $m_{\textrm{hy}}\ll M_{\mathrm{T}} \lesssim M_{\mathrm{N}} $ as
\begin{subequations}
\label{mass1}
\begin{eqnarray}
M_{1}&=&M_{\mathrm{T}} -\frac{M_{\mathrm{T}} +\mathrm{\cos }(2\alpha )M_{\mathrm{N}} }{M_{\mathrm{N}} ^{2}-M_{\mathrm{T}} ^{2}%
}~m_{\textrm{hy}}^{2}+\mathcal{O}\left(\frac{m_{\textrm{hy}}^4}{M_{\mathrm{N,T}}^4}\right), \ \ \ \ \ \ \  \\
M_{2}&=&M_{\mathrm{N}} +\frac{\mathrm{\cos }(2\alpha )M_{\mathrm{T}} +M_{\mathrm{N}} }{M_{\mathrm{N}} ^{2}-M_{\mathrm{T}} ^{2}%
}~m_{\textrm{hy}}^{2}+\mathcal{O}\left(\frac{m_{\textrm{hy}}^4}{M_{\mathrm{N,T}}^4}\right).
\end{eqnarray}
\end{subequations}
If we use $\mathbf{T}.\mathbf{D}^{2}=\mathbf{H.}\mathbf{T}$ we find the following relations between the two pairs of parameters:
\begin{eqnarray}
\tan \phi &\simeq &\frac{M_{\mathrm{N}} -M_{\mathrm{T}} }{M_{\mathrm{N}} +M_{\mathrm{T}} }\tan \alpha, \\
\tan \theta &\simeq &\frac{ M_{\mathrm{N}} ^{2}+2M_{\mathrm{N}} M_{\mathrm{T}} \cos 2\alpha
+M_{\mathrm{T}} ^{2}}{ M_{\mathrm{T}} ^{2}-M_{\mathrm{N}} ^{2} } \nonumber\\
&\times& \frac{m_{\textrm{hy}} }{  M_{\mathrm{N}} \cos
\left( \alpha -\phi \right) +M_{\mathrm{T}} \cos \left( \alpha +\phi \right)}
\end{eqnarray}%
\newline
The physical eigenstates are now given by the application of $\mathbf{T}$ in Eq.~(\ref{Trans}) to the symmetry states, $\mathbf{T.}N_{R}$,
\begin{equation}
\left(
\begin{array}{c}
N_{1R} \\
N_{2R}%
\end{array}%
\right) =\left(
\begin{array}{c}
N_{R}\cos \theta +T_{R}^{0}e^{-i\phi }\sin \theta \\
T_{R}^{0}\cos \theta -N_{R}e^{i\phi }\sin \theta%
\end{array}%
\right) .
\end{equation}

\begin{figure}[th]
\centering
\includegraphics[height=0.6\columnwidth]{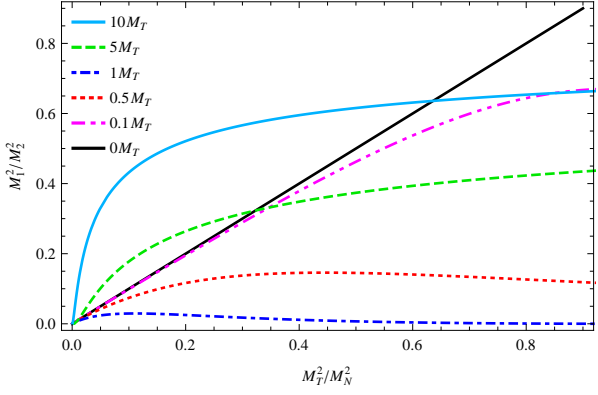}
\caption{Relation between the mass ratios $M_{1}^{2}/M_{2}^{2}$ and $M_{\mathrm{T}} ^{2}/M_{\mathrm{N}} ^{2}$ for numerical diagonalization for $0<m_{\textrm{hy}}<10M_{\mathrm{T}} $.}
\label{figMASSNcomb}
\end{figure}

Now, by using  Eq.~(\ref{mass1}) it is easy to show that we can get $M_{\mathrm{N}} $
and $M_{\mathrm{T}} $ written in the mass eigenstate basis by the following nonlinear transformation up to the order $\mathcal{O}{(m_\textrm{hy}^2/{M_{\mathrm{N,T}}^2})}$:
\begin{subequations}
\label{MNT}
\begin{eqnarray}
M_{\mathrm{T}} &\simeq&M_{1}\left( 1+\frac{m_{\textrm{hy}}^{2}}{M_{2}^{2}-M_{1}^{2}}\right)
+M_{2}\left( \frac{m_{\textrm{hy}}^{2}~\mathrm{\cos }2\alpha}{M_{2}^{2}-M_{1}^{2}}%
\right) , \ \ \ \ \ \ \   \\
M_{\mathrm{N}} &\simeq&M_{2}\left( 1-\frac{m_{\textrm{hy}}^{2}}{%
M_{2}^{2}-M_{1}^{2}}\right)-M_{1}\left( \frac{m_{\textrm{hy}}^{2}~\mathrm{\cos }2\alpha}{%
M_{2}^{2}-M_{1}^{2}}\right). \ \ \ \ \ \ \
\end{eqnarray}
\end{subequations}
We notice that these approximations are in great accordance for values of $m_\textrm{hy}\lesssim 0.1M_1$. It should also be emphasized that the hybrid seesaw I+III only takes place when $\Sigma$ gets a VEV. Without the symmetry breaking by $\Sigma$, the model has diagonal heavy masses, but there are still new contributions to CP violation due to interferences of tree- and one-loop-level diagrams, as we will see below. The exact numerical relation between the ratios $M_{1}^{2}/M_{2}^{2}$ and $M_{\mathrm{T}} ^{2}/M_{\mathrm{N}} ^{2}$ is shown in Fig.~\ref{figMASSNcomb} for $m_{\textrm{hy}}$ in the range $0<m_{\textrm{hy}}<10M_{\mathrm{T}} $. Deviations from a straight line when $m_{\textrm{hy}}\neq 0$ shows that the CP violation may have an unexpected behavior in terms of $M_{1}^{2}/M_{2}^{2}$ since it may not be exactly written as a bijective function of $M_{1}^{2}/M_{2}^{2}$, giving rise to complications in the broken phase. Notice also that $M_{1}^{2}/M_{2}^{2}$ is very small for $m_{\textrm{hy}}\simeq 1 M_{\mathrm{T}}$.

\section{CP Violation in the $1N1T$ Setup}
\label{sec:CP1N1T}

\begin{figure}[h!]
\minipage{0.15\textwidth}
\subfloat[\label{tree0}]{\includegraphics[height=0.63\columnwidth]{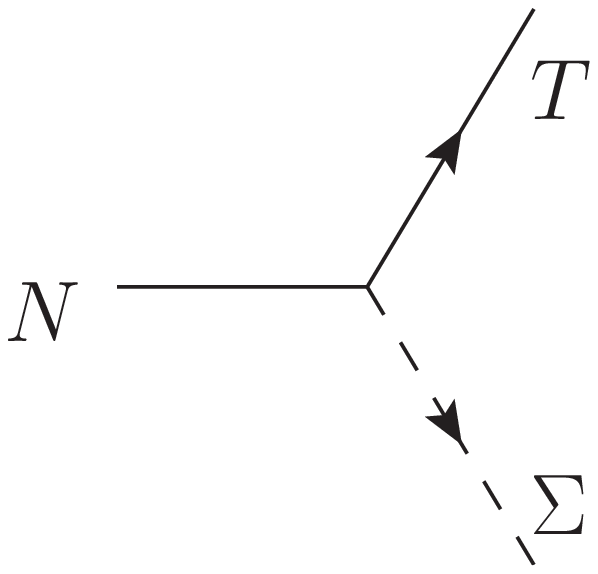}}
\endminipage
\hfill
\minipage{0.15\textwidth}
\subfloat[\label{tree0_0}]{\includegraphics[height=0.63\columnwidth]{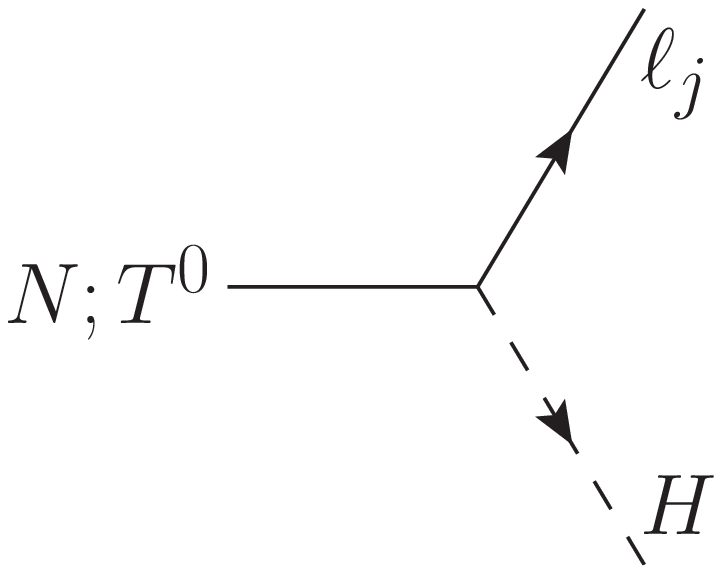}}
\endminipage
\hfill
\minipage{0.15\textwidth}
\subfloat[\label{vertex0}]{
\includegraphics[height=0.63\columnwidth]{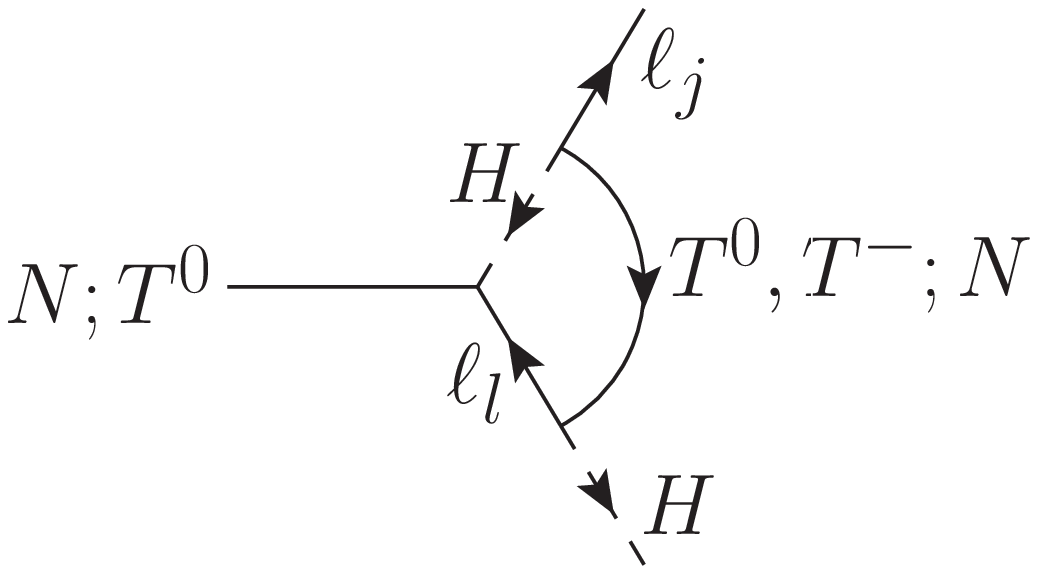}}
\endminipage
\hfill
\minipage{0.15\textwidth}
\subfloat[\label{self0}]{\includegraphics[height=0.63\columnwidth]{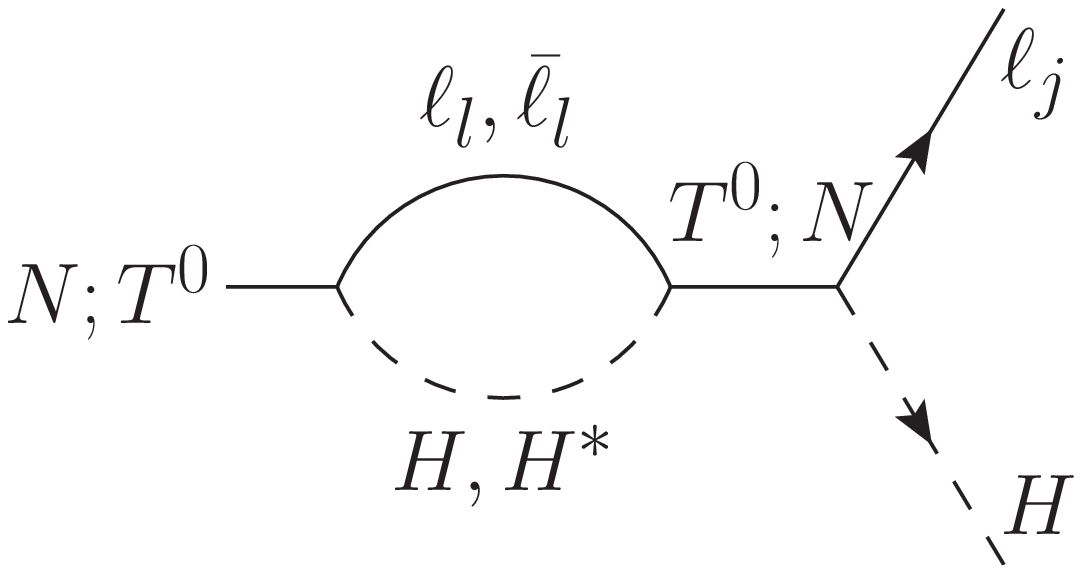}}
\endminipage
\hfill
\caption{Tree-level and one-loop-level contributions to CP asymmetry in the $N$ and $T$ decays with $\ell H$ and $T\Sigma$ final states in the minimal 1N1T model.}
\label{fig:mintree}
\label{fig:minloop}
\end{figure}

Let us assume that the Universe is in the symmetric phase, specifically when $\Sigma$ has not yet acquired a VEV, and therefore $\mathbf{M}$ is diagonal and real in this WB. Let us assume that $T^{0}$ is always lighter than the singlet $N$ and their
decay channels at tree level are described in the diagrams showed in Fig.~\ref{fig:mintree}. The CP asymmetry is generally defined by~\cite{Branco:2011zb}
\begin{equation}
\varepsilon _{j(fS) }^{(\mathrm{N})}=\frac{\Gamma _{\left( N\rightarrow
f_{j}S\right) }-\Gamma _{\left( N\rightarrow \bar{f}_{j}S^\dagger%
\right) }}{\Gamma _{\left( N\rightarrow fS\right) }+\Gamma _{\left(
N\rightarrow \bar{f}S^\dagger\right) }},  \label{eps}
\end{equation}%
where $\Gamma _{\left( N\rightarrow fS\right) }\equiv \sum_{j=all}\Gamma
_{\left( N\rightarrow f_{j}S\right) }$ is the total decay width of $N$.
A similar expression gives the $CP$ asymmetry in $T^0$ decays. Notice that the singlet $N$ can also decay to a $T\Sigma$ final state. Thus, a kind of sequential decay mechanism can be applied here and the decay of $N$ should not be erased by the late $T^0$ decays, as in the $N_2$-dominated scenario~\cite{DiBari:2005st}.

Nonvanishing CP asymmetry arises from the interferences of tree-level diagrams in Fig.~\ref{tree0_0} with their respective one-loop diagrams in Figs.~\ref{vertex0} and ~\ref{self0}~\cite%
{Covi:1996wh,Davidson:2008bu}. In the present model the CP violation in the $%
N$ decays reads as
\begin{equation}
\varepsilon ^{\left( N\right) }=\frac{\textrm{Im}\left[ \left( y^{t\dagger
}y^{\nu }\right) ^{2}\right] }{8\pi\left( y^{\nu \dagger }y^{\nu
}\right) +12\pi \mathbb{Y}(w,y^{_\Sigma}) }\left( \frac{\sqrt{w}}{1-w}+3f\left( w\right)
\right),  \label{e_minN}
\end{equation}%
where $w\equiv M_{\mathrm{T}} ^{2}/M_{\mathrm{N}} ^{2}$ (notice that $w<1$ due to the inversion of the usual definition). We have summed over the final states (vanilla leptogenesis) to elucidate the properties of this mechanism. Here $f(x)$ is the usual vertex one-loop function defined by
\begin{equation}
f\left( x\right) =\sqrt{x}\left[ 1-\left( 1+x\right) \ln \left( 1+\frac{1}{x}%
\right) \right].\label{eq:func1loop}
\end{equation}%
We have also defined the function $\mathbb{Y}$ as
\begin{equation}
\mathbb{Y}(w,y^{_\Sigma})=\left( w+1\right) \left\vert y^{{_\Sigma}}\right\vert ^{2}+%
\sqrt{w}\left( \left( y^{_\Sigma \ast }\right) ^{2}+\left( y^{_\Sigma }\right)
^{2}\right),\label{eq:defY}
\end{equation}
which means $\mathbb{Y}(w,0)=0$ in the decoupling limit of both seesaw mechanisms. Notice that Fig.~\ref{tree0} does not have a nonvanishing one-loop counterpart, and thus it does not contribute to CP violation in 1N1T case, although it contributes to the total decay width, as can be seen in the denominator of Eq.~(\ref{e_minN}).

The CP violation generated in $T^{0}$ decays is given by
\begin{equation}
\varepsilon ^{\left( T\right) }=\frac{\textrm{Im}\left[ \left( y^{t\dagger
}y^{\nu }\right) ^{2}\right] }{8\pi \left( y^{t\dagger }y^{t}\right) }\left(
\frac{\sqrt{w}}{1-w}-f\left( \frac{1}{w}\right) \right).  \label{e_minT}
\end{equation}%
 If we take the limit for the lightest heavy state (triplet) ($w\ll 1$) in Eq.$( \ref{e_minT})$ the well-known result is recovered,
\begin{equation}
\varepsilon ^{\left( T\right) }\overset{M_{\mathrm{T}} \ll M_{\mathrm{N}} }{\simeq }\frac{3\sqrt{w}}{%
16\pi }\frac{\textrm{Im}\left[ \left( y^{t\dagger }y^{\nu
}\right) ^{2}\right] }{\left( y^{t\dagger }y^{t}\right) }.  \label{eT}
\end{equation}%
In this case the CP asymmetry generated in $N$ decays can be neglected, and we recover the usual approximation when the lightest heavy state dominates the leptogenesis. The difference in this case concerns only the neutrino
mass generation when $\Sigma$ gets a VEV.

To get some insight about the CP asymmetries, let us consider the following simplifications,
\begin{equation}\label{eq:case1}
y^t=\mathcal{O}(y_t), \,\,  y^\nu = \mathcal{O}(y_\nu)(1+i), \,\, \frac{m_\textrm{hy}}{v_{_\Sigma}}\sim \mathcal{O}(y_\nu).
\end{equation}
For $M^2_{\mathrm{T}}/ M^2_{N}=w \sim 10^{-3}$ and $\mathcal{O}(y)'s\sim 10^{-3}$, we get $\varepsilon ^{\left( T\right) }\sim 10^{-10}$
while $\varepsilon ^{\left( N\right) }$ should be almost erased or negligible~\cite{Buchmuller:2004nz}. In this
decoupled limit there is no dependence on the $y^{_\Sigma }$ coupling; then, the
low-energy connection is almost the same as the standard type-I or type-III
leptogenesis mechanisms. On the other hand, as the masses get closer the $\varepsilon^N$ is no longer negligible and can be higher than $\varepsilon^T$ for $\mathcal{O}(y_\nu)\lesssim 10^{-1}\mathcal{O}(y_t)$.

From Eq.~(\ref{eq:vsig}) we can conclude that $y^{_\Sigma}=\sqrt{2}(m_\textrm{hy}/v_{_\Sigma})e^{i\alpha}$, and thus using Eq.~(\ref{eq:case1}) we can rewrite $\mathbb{Y}$ as
\begin{equation}\label{Yw}
  \mathbb{Y}=2\left(\frac{m_\textrm{hy}}{v_{_\Sigma}}\right)^2\left[1+w+2\sqrt{w}\cos{2\alpha} \right].
\end{equation}
Using the above equation, one can rewrite Eqs.~(\ref{e_minN}) and~(\ref{e_minT}) as
%
\begin{equation}\label{eq:epsw}
  \varepsilon^{(N)}(w,\alpha)\simeq\frac{\left(\frac{\sqrt{w}}{1-w}+3f(w)\right)}{4\pi\left[5+3w+6\sqrt{w}\cos{2\alpha} \right]}\mathcal{O}(y_t^2),
\end{equation}
and
\begin{equation}\label{eq:epswT}
  \varepsilon^{(T)}(w)\simeq\frac{1}{4\pi}\left(\frac{\sqrt{w}}{1-w}-f\left(\frac{1}{w}\right)\right)\mathcal{O}(y_\nu^2),
\end{equation}
respectively.
\begin{figure}[th]
\centering
\includegraphics[height=0.6\columnwidth]{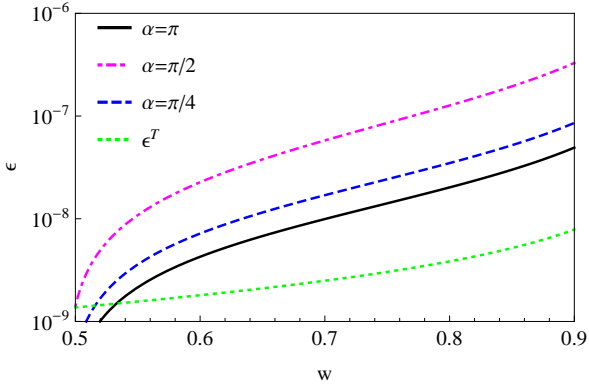}
\caption{Numerical CP violation in N and T decays [Eq.~(\ref{eq:epsw}) and Eq.~(\ref{eq:epswT})] in terms of $w=M_{\mathrm{T}}^2/M_{\mathrm{N}}^2$. The dotted (green) line is the $\varepsilon^T$, while the others are values of $\varepsilon^N$ for $\alpha=\pi/4,\pi/2,\pi$, $O(y_\nu)= 10^{-4}$, and $O(y_t)= 10^{-3}$.}
\label{figeps1}
\end{figure}
In Fig.~\ref{figeps1} we can see an example of the suppression of $\varepsilon^T$ compared to $\varepsilon^N$ for
the results presented in Eqs.~(\ref{eq:epsw}) and~(\ref{eq:epswT}), shown in the range $0.5\le w\le 0.9$, with $O(y_\nu)= 10^{-4}$ and $O(y_t)= 10^{-3}$. The study of a full set of Boltzmann equations could be necessary to investigate the baryon-to-photon ratio behavior since the corresponding CP asymmetries are generated displaced for a very small $w$ ($w\lesssim 0.5$). We will not consider this possibility in this simplified example once $0.71 M_{\mathrm{N}} \lesssim M_{\mathrm{T}} \lesssim 0.95 M_{\mathrm{N}}$ to avoid an extremely suppressed $\varepsilon^{(N)}$.

\section{CP Violation in the $2N1T$ Setup}

\label{sec:CP2N1T}
The next extension (with two fermion singlets
and one fermion triplet) has the most general Lagrangian, which is similar to the one presented in Eq.~(\ref{eq:LagMin}) but with the following small changes: $y_{i}^\nu\rightarrow y_{im}^\nu, y^{_\Sigma}\rightarrow y_{m}^{_\Sigma}, M_{\mathrm{N}}\rightarrow M_{{\mathrm{N}_m}}$ (all real), and $N\rightarrow N_m$.
We assume also that $\Sigma$ has not yet acquired a VEV and the model starts with a diagonal $\mathbf{M}$ without the need for writing the broken mass eigenstates.

\begin{figure}[h!]
\minipage{0.15\textwidth}
\subfloat[\label{self1}]{\includegraphics[height=0.63\columnwidth]{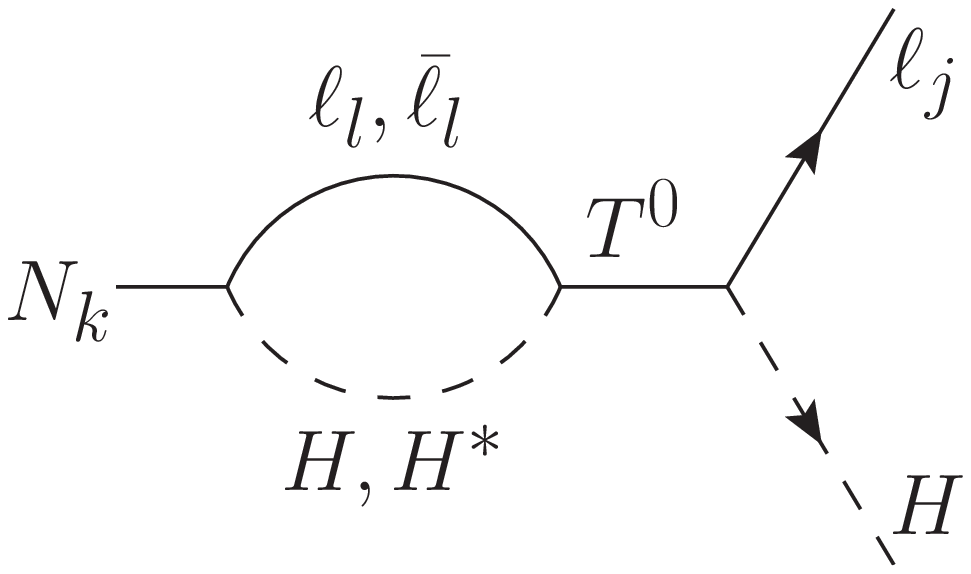}}
\endminipage
\hfill
\minipage{0.15\textwidth}
\subfloat[\label{vertex1}]{\includegraphics[height=0.63\columnwidth]{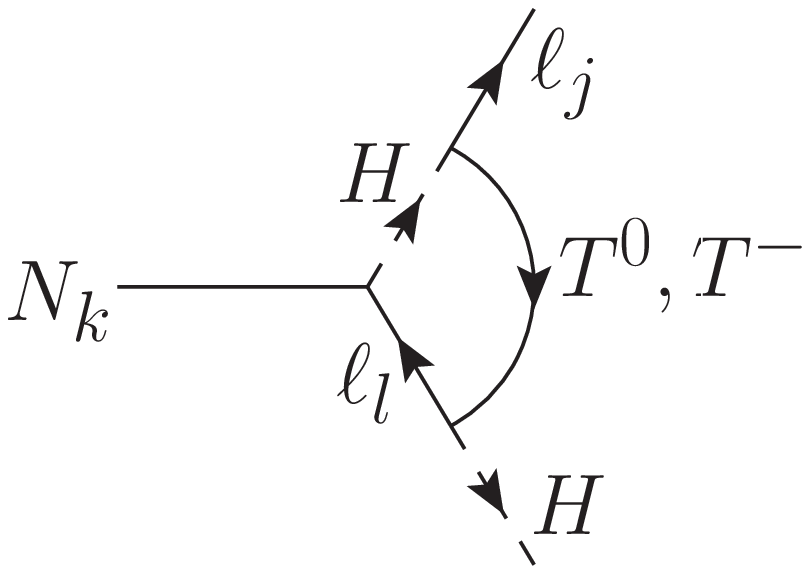}}
\endminipage
\hfill
\minipage{0.15\textwidth}
\subfloat[\label{self2}]{\includegraphics[height=0.63\columnwidth]{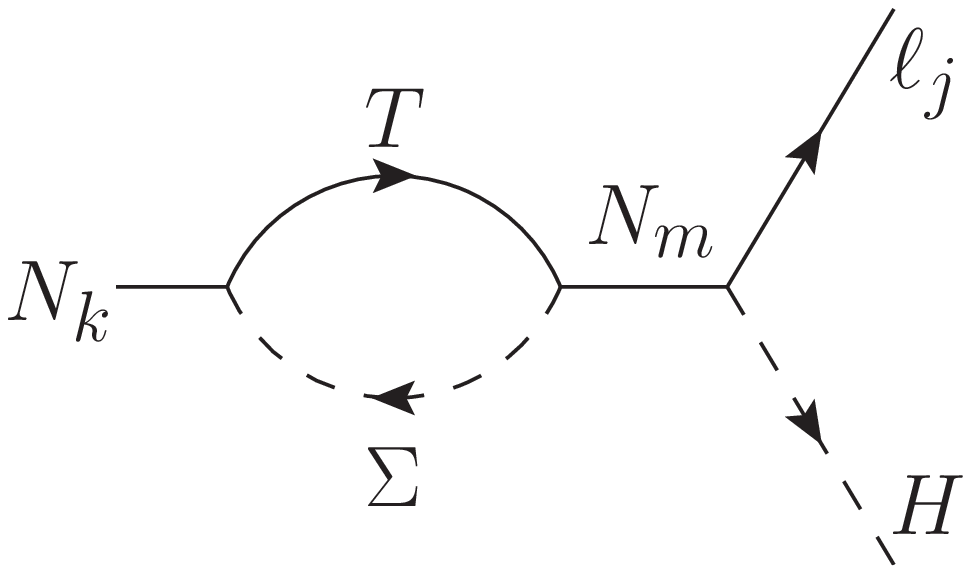}}
\endminipage
\hfill
\caption{One-loop-level diagrams contributing to CP asymmetry in the $N_k$ decays with triplet components in the propagators.}\label{1loopTprop}
\end{figure}

\begin{figure}[h!]
 \minipage{0.15\textwidth}
\subfloat[\label{tree1}]{\includegraphics[height=0.63\columnwidth]{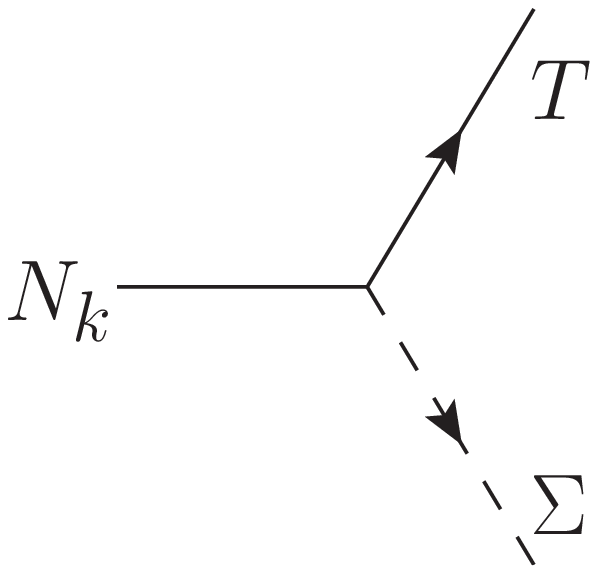}}
\endminipage
\hfill
\minipage{0.15\textwidth}
\subfloat[\label{self3}]{\includegraphics[height=0.63\columnwidth]{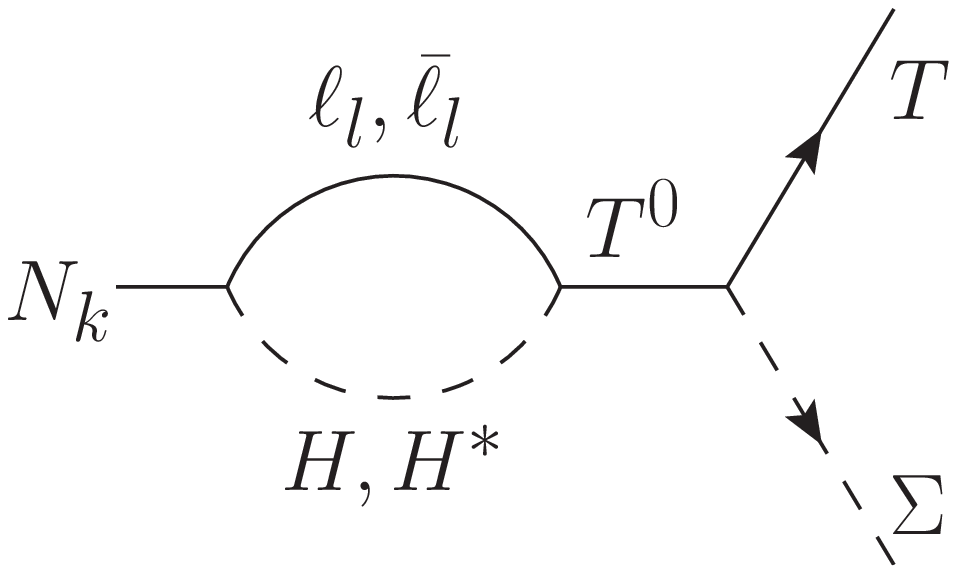}}
\endminipage
\hfill
\minipage{0.15\textwidth}
\subfloat[\label{self4}]{\includegraphics[height=0.63\columnwidth]{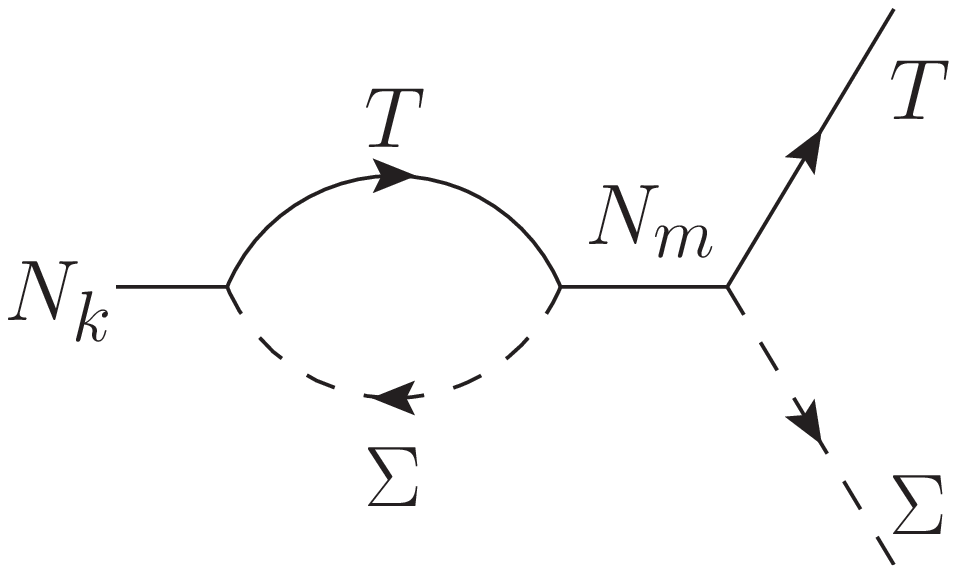}}
\endminipage
\hfill
\minipage{0.15\textwidth}
\subfloat[\label{vertex2}]{\includegraphics[height=0.63\columnwidth]{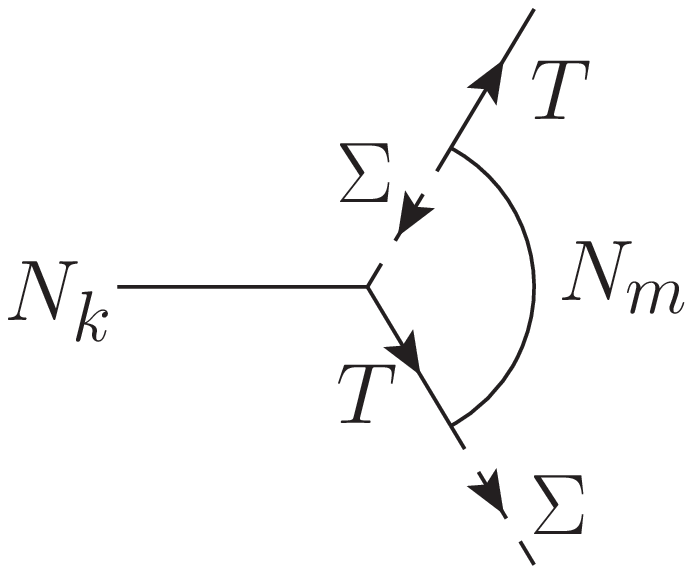}}
\endminipage
\hfill
\caption{Tree-level and one-loop-level contributing to CP asymmetry in the $N_k$ decays with a $T\Sigma$ final state.}\label{1loopTfinal}
\end{figure}

For the study of CP violation let us concentrate on the decays of
the lightest heavy Majorana singlet, $N_{k}$, and also on the neutral
fermion triplet decays, as before. The CP violation generated by the $N_{k}$ decays [Eq. $(\ref{eps})$] is $\varepsilon _{k,j\left( fS\right) }$. $\Gamma _{\left( N_{k}\rightarrow fS\right) }\equiv \sum_{j=all}\Gamma
_{\left( N_{k}\rightarrow f_{j}S\right) }$ is the total decay width of the $%
N_{k}$.
The first crucial difference between the $2N1T$ and $1N1T$ schemes is in the fermion propagator with heavy triplets inside the loop and the new final-state channel $T\Sigma $ (also at
one-loop level), as illustrated in Fig.~\ref{1loopTprop} and Fig.~\ref{1loopTfinal}, respectively, which increase the total CP asymmetry.

The $N_{k}$ total width at tree-level is now settled by $\Gamma _{\left( N_{k}\rightarrow
all\right) }=\Gamma _{\left( N_{k}\rightarrow \ell H\right) }+\Gamma
_{\left( N_{k}\rightarrow \bar{\ell}H^{\dagger}\right) }+\Gamma _{\left(
N_{k}\rightarrow T\Sigma \right) }+\Gamma _{\left( N_{k}\rightarrow
\bar{T}\Sigma ^{\dagger}\right) }$ and is explicitly given by
\begin{eqnarray}
\Gamma _{\left( N_{k}\rightarrow all\right) }&=&\frac{M_{_{\mathrm{N}k}}}{8\pi }\left(
y^{\nu \dagger }y^{\nu }\right) _{kk}+\frac{3}{16\pi }\bigg[ \left( M_{_{\mathrm{N}k}}+%
\frac{M_{\mathrm{T}} ^{2}}{M_{_{\mathrm{N}k}}}\right) \left\vert y_{k}^{_\Sigma }\right\vert
^{2} \nonumber \\ &&+ M_{\mathrm{T}} \left[ \left( y_{k}^{_\Sigma \ast }\right) ^{2}+\left(
y_{k}^{_\Sigma }\right) ^{2}\right] \bigg] ,  \label{GamTot}
\end{eqnarray}%
where we have used the fact that $\Gamma _{\left( N_{k}\rightarrow \ell H\right)
}=\Gamma _{\left( N_{k}\rightarrow \bar{\ell}H^{\dagger}\right) }$ and $\Gamma
_{\left( N_{k}\rightarrow T\Sigma \right) }=\Gamma _{\left( N_{k}\rightarrow
\bar{T}\Sigma^\dagger\right) }$ at lowest order.

\begin{figure}[h!]
\minipage{0.15\textwidth}
\subfloat[\label{SM0}]{\includegraphics[height=0.63\columnwidth]{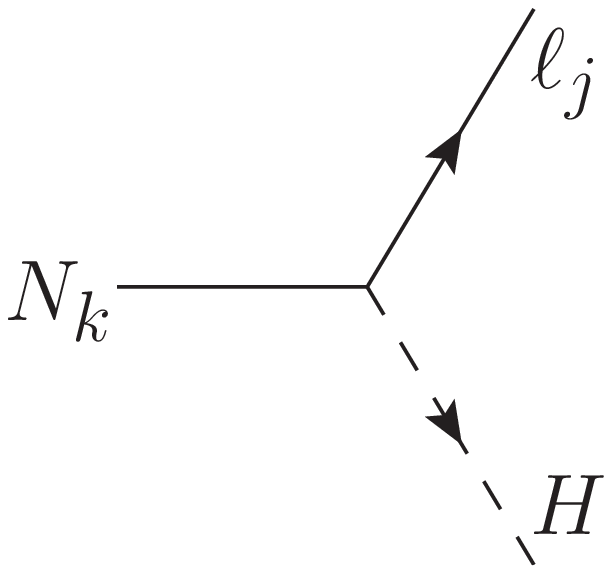}}
\endminipage
\hfill
\minipage{0.15\textwidth}
\subfloat[\label{SMself}]{\includegraphics[height=0.63\columnwidth]{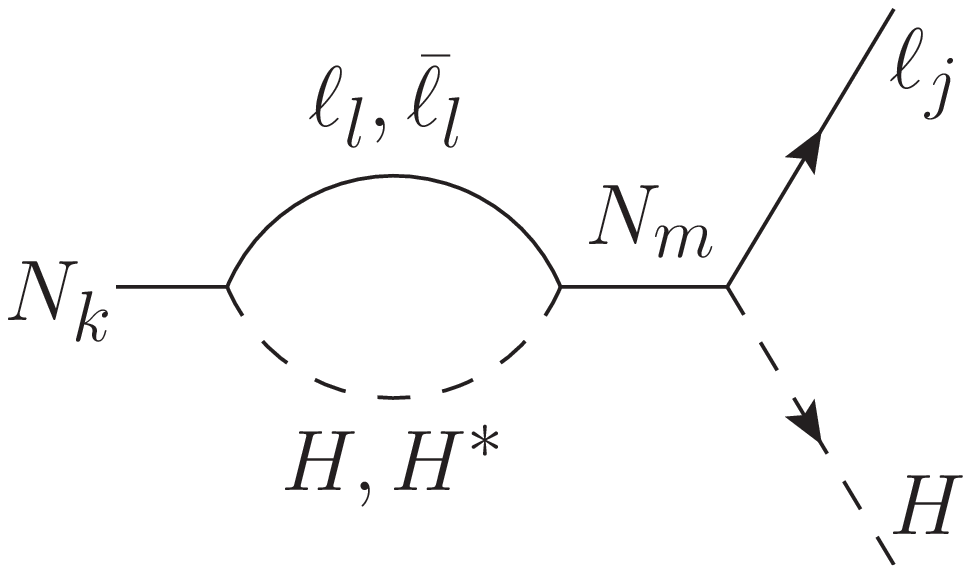}}
\endminipage
\hfill
\minipage{0.15\textwidth}
\subfloat[\label{SMvertex}]{\includegraphics[height=0.63\columnwidth]{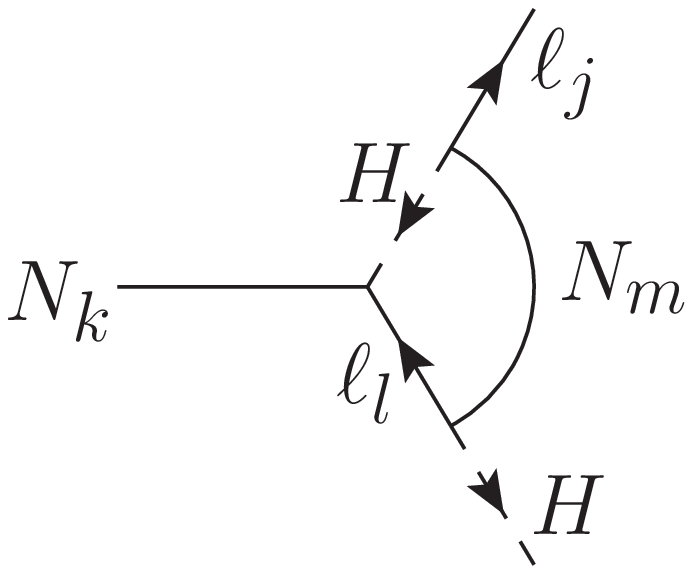}}
\endminipage
\hfill
\caption{Tree-level and one-loop-level diagrams contributing to CP asymmetry in the $N_k$ decays with a $\ell H$ final state.}\label{fig:2NtoHL}
\end{figure}

The usual CP violation in the decay of $N_k$ is due to the interference
of the diagrams in Fig.~\ref{SM0} with Figs.~\ref{SMself} and ~\ref{SMvertex}, which yields
\begin{eqnarray}
\varepsilon _{k,j}^{\left( SM\right) }&=&-\frac{
1 }
 {8\pi\left( y^{\nu \dagger }y^{\nu
}\right) _{kk}
+12\pi\mathbb{Y}\left(w_k,y_k^{_\Sigma} \right)
  } \times
  \nonumber \\ &&
 \sum_{m\neq k}\textrm{Im}\bigg\{y_{kj}^{\nu \dagger }y_{jm}^{\nu }
\bigg[ \left( y^{\nu \dagger }y^{\nu
}\right) _{km}\left( \frac{\sqrt{x_{k}^{m}}}{1-x_{k}^{m}}+f\left(
x_{k}^{m}\right) \right)  \nonumber \\ &&
+\left( y^{\nu \dagger }y^{\nu }\right) _{mk}\frac{1%
}{1-x_{k}^{m}}\bigg] \bigg\},
\label{eps_sm}
\end{eqnarray}%
and the modification in the denominator comes from the total width decay, given by Eq.~(\ref{GamTot}).
Here $x_{k}^{m}\equiv M_{_{\mathrm{N}m}}^{2}/M_{_{\mathrm{N}k}}^{2}$ and $w_{k}\equiv
M_{\mathrm{T}} ^{2}/M_{_{\mathrm{N}k}}^{2}$. The CP asymmetry given by the interference
of the tree level in Fig.~\ref{SM0} with the one-loop levels in Figs.~\ref{self1} and ~\ref{vertex1}
is $\varepsilon _{kj}^{\left( 1\right) }=\varepsilon^{(1)} _{\textrm{self}-k,j}+\varepsilon^{(1)}
_{\textrm{vert}-k,j}$ (respectively) and reads as
\begin{eqnarray}
\varepsilon _{kj}^{\left( 1\right) }&=&-\frac{
1 }
 {8\pi\left( y^{\nu \dagger }y^{\nu
}\right) _{kk}
+12\pi\mathbb{Y}\left(w_k,y_k^{_\Sigma} \right)
  } \times
  \nonumber \\ &&
    \sum_{m\neq k}\textrm{Im}\bigg\{
y_{kj}^{\nu \dagger }y_{j}^{t}\bigg[ \left( y^{\nu \dagger }y^{t}\right)
_{k}\left( \frac{\sqrt{w_{k}}}{1-w_{k}}
+3f\left( w_{k}\right) \right)  \nonumber \\ &&
+\left( y^{t\dagger }y^{\nu }\right) _{k}\frac{1}{1-w_{k}}
\bigg]
\bigg\},
\label{eps_sv1}
\end{eqnarray}%
 which corresponds to a permutation of $%
N_{m}$ and $T$ $\ $ in Eq.~(\ref{eps_sm}) with suitable
consideration of a factor $\sqrt{2}$ in the vertex with charged triplet components. The interference of the diagram in Fig.~\ref{SM0} and in Fig.~\ref{self2} gives the new source of CP violation,
\begin{eqnarray}
\varepsilon _{k,j}^{\left( 2\right) }~&=&\frac{3}{16\pi\left( y^{\nu \dagger }y^{\nu
}\right)_{kk} +24\pi \mathbb{Y}(w_k,y_k^{_\Sigma})  }\sum_{m\neq k}\frac{%
\left( 1-w_{k}\right) }{\left( 1-x_{k}^{m}\right) }
\times
\nonumber \\
&& \textrm{Im}
\bigg\{
y_{kj}^{\nu \dagger }y_{jm}^{\nu }
\bigg[
y_{k}^{_\Sigma }\left( \left(
w_{k}+1\right) y_{m}^{_\Sigma \ast }-2\sqrt{x_{k}^{m}w_{k}}y_{m}^{_\Sigma
}\right)
\nonumber \\
&&
+y_{k}^{_\Sigma \ast }\left( \sqrt{x_{k}^{m}}\left( w_{k}+1\right)
y_{m}^{_\Sigma }-2\sqrt{w_{k}}y_{m}^{_\Sigma \ast }\right)
\bigg]
\bigg\},
 \label{eps_2}
\end{eqnarray}
and it should be written explicitly as a function of $\alpha_i$ since the latter are the new CP-violating phases.
Notice that we have not summed over the final flavors $j$ in the results
given in Eqs.~(\ref{eps_sv1}) and~(\ref{eps_2}). These equations hold even for more than two singlets. However,
these equations change if we introduce two or more triplets with two singlets
and we will not address this general setup.

The CP contribution that comes from
the interference of the tree-level in Fig.~\ref{tree1} with the one-loop levels in Figs.~\ref{self3},~\ref{self4}, and~\ref{vertex2} vanishes since only one triplet has been taken into account. The diagram in Fig.~\ref{self3}\
automatically cancels out the imaginary combination of couplings when
it is summed over the lepton-flavor-conserving and -violating loop diagrams.
Each of the diagrams in Figs.~\ref{self4} and~\ref{vertex2} has a vanishing imaginary
interference combination of couplings with the tree-level diagram in Fig.~\ref{tree1} and do not contribute to CP asymmetry with only one triplet.


\label{sec:CP2N1TbyT}
  \begin{figure}[h!]
\minipage{0.15\textwidth}
\subfloat[\label{treeT}]{\includegraphics[height=0.63\columnwidth]{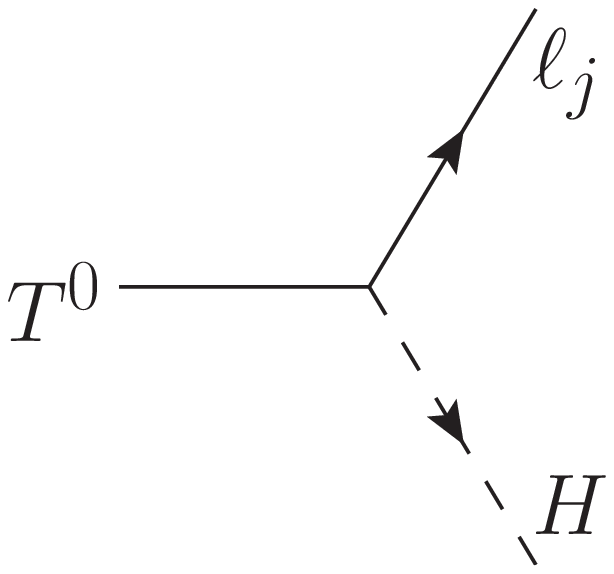}}
\endminipage
\hfill
\minipage{0.15\textwidth}
\subfloat[\label{selfT}]{\includegraphics[height=0.63\columnwidth]{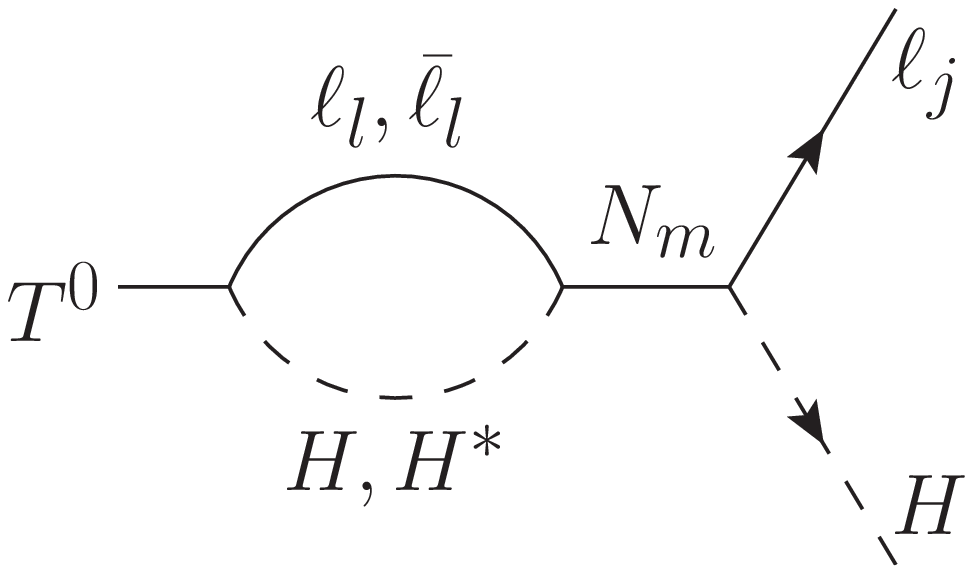}}
\endminipage
\hfill
\minipage{0.15\textwidth}
\subfloat[\label{vertexT}]{\includegraphics[height=0.63\columnwidth]{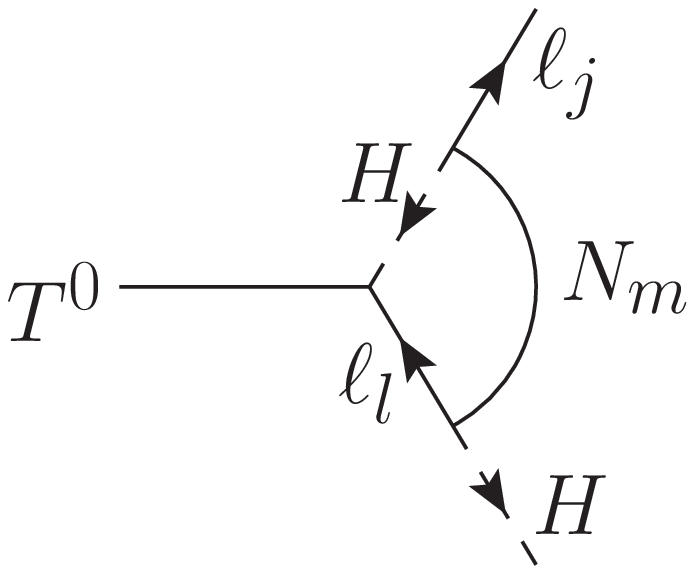}}
\endminipage
\hfill
\caption{Tree-level and one-loop-level contributing to CP asymmetry in the $N_k$ decays with a $\ell H$ final state.}\label{fig:animals}
\end{figure}

The CP violation in the interference of the tree-level diagram in Fig.~\ref{treeT} with the one-loop levels in Fig.~\ref{selfT} and~\ref{vertexT} leads to%
\begin{eqnarray}
\varepsilon _{j}^{\left( T\right) }&=&-\frac{1}{8\pi \left( y^{t\dagger
}y^{t}\right) }\sum_{m}\textrm{Im}
\bigg\{
y_{j}^{t\ast }y_{jm}^{\nu }\bigg[
\left( y^{t\dagger }y^{\nu }\right) _{m}
\bigg( \frac{\sqrt{w_{m}}}{w_{m}-1}\nonumber%
\\&&+
f\bigg( \frac{1}{w_{m}}\bigg) \bigg)
 +\left( y^{\nu \dagger }y^{t}
\right)
_{m}\frac{w_{m}}{w_{m}-1}
\bigg]
\bigg\}.
\end{eqnarray}

\begin{figure}[th]
\centering
\includegraphics[height=0.6\columnwidth]{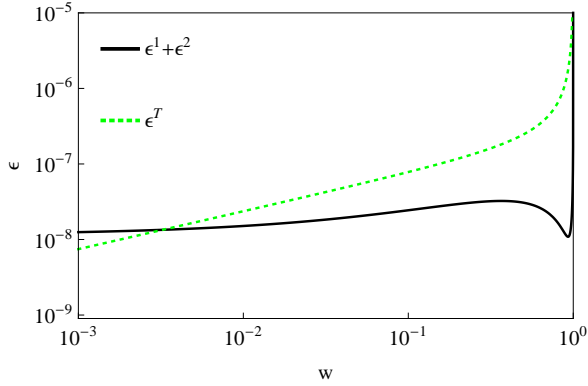}
\caption{Numerical CP violation in $N_1$ and $T$ decays [Eq.~(\ref{e1apA})-~(\ref{eTap})] in terms of $w=M_{\mathrm{T}}^2/M_{\mathrm{N}_1}^2$ for $\alpha_1=\pi/2$, $\alpha_2=0$, $x=M^2_{\mathrm{N}_2}/M^2_{\mathrm{N}_1}=10$ and $\mathcal{O}(y_\nu)=10^{-3}$, $\mathcal{O}(y_t)=10^{-4}$.}
\label{figeps3}
\end{figure}

For simplification, let us assume the hierarchical order $N_1$, $T^0$, $N_2$, from the lightest to the heaviest, to give some results in the unflavored regime (summing over the final flavors), considering only the rough order of couplings $y^{\nu}$, $y^{t}$ and $y^{_\Sigma}$. If we consider the crude generalization of Eq.~(\ref{eq:case1}) for all matrix elements,
\begin{equation}\label{eq:case2}
y_k^t=\mathcal{O}(y_t), \,\,  y_{k1,2}^\nu = \mathcal{O}(y_\nu)(1+i), \,\, y^{_\Sigma}_{k}\sim \sqrt{2}e^{i\alpha_k}\mathcal{O}(y_\nu),
\end{equation}
then we get $\varepsilon _{1}^{\left( SM\right) }=0$. For a realistic low-energy connection it should also be related to $\sin \theta_{13}$ in general~\cite{Branco:2011zb}, but here the condition for (flavored) leptogenesis $\sin \theta_{13}\gtrsim 0.09$~\cite{Branco:2006ce} does not necessarily take place. Thus, the simplified expressions for CP violation are roughly given by
\begin{equation}
\varepsilon _{1}^{(1)}\simeq \frac{\left( \frac{\sqrt{w}}{1-w}+3f\left(
w\right) \right) }{4\pi \left( 3+w+2\sqrt{w}\cos 2\alpha _{1}\right) }%
O\left( y_{t}^{2}\right)\label{e1apA}
\end{equation}
and
\begin{eqnarray}
\varepsilon _{1}^{\left( 2\right) } &\simeq& \frac{1}{4\pi \left( 3+w+2\sqrt{%
w}\cos 2\alpha \right) }
\bigg[\frac{\left( 1-w^2\right) }{\left( 1+\sqrt{x}\right) } \sin \left( \alpha _{1}-\alpha _{2}\right)\nonumber \\&+&2%
\frac{\left( 1-w\right) }{\left( 1+\sqrt{x}\right) }\sqrt{w}\sin \left( \alpha _{1}+\alpha _{2}\right) \bigg] O\left( y_{\nu
}^{2}\right).\label{e2apA}
\end{eqnarray}%
Finally, the triplet gives rise to the following CP asymmetry:
\begin{equation}
\varepsilon ^{(T)}\simeq -\frac{3}{8\pi }\left[ \frac{\sqrt{w}}{%
w-1}+f\left( \frac{1}{w}\right) +\frac{\sqrt{wx}}{w-x}+f\left( \frac{x}{w}%
\right) \right] O\left( y_{\nu}^{2}\right),\label{eTap}
\end{equation}
where we have unequivocally denoted $w\equiv w_1$ and $x\equiv x_1^2$. The final CP violation in the $N_1$ decays is given by $\varepsilon_1 =\varepsilon _{1}^{\left(
SM\right) }+\varepsilon _{1}^{\left( 1\right) }+\varepsilon _{1}^{\left(
2\right) }$. If $\varepsilon^{(T)}$ is suppressed or vanishes by decaying in the equilibrium stage~\cite{Sakharov:1967dj},
the $N_1$ drives the leptogenesis mechanism. Our numerical results are shown in Fig.~\ref{figeps3} for $x=M^2_{\mathrm{N}_2}/M^2_{{\mathrm{N}_1}}=10$. The naive approximations shown in Eqs.(\ref{e1apA})~-~(\ref{eTap}) may get corrections if we proceed to a flavored analysis~\cite{flavored}. It is straightforward to verify that
if only one singlet and two triplets are taking into account the result is
still almost the same with minimal interchanges of couplings. On the other hand, if two singlets plus two triplets are included in the model, the results
will change drastically due to the fact that nonvanishing contributions from the graphs in Figs.~\ref{self3}-~\ref{vertex2} are not canceled out and may not be suppressed.

\begin{figure}[th]
\centering
\includegraphics[height=0.6\columnwidth]{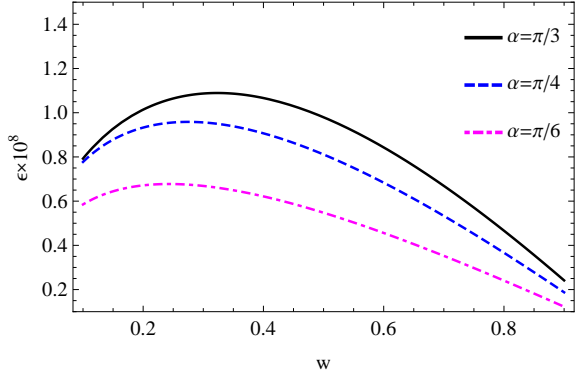}
\caption{Numerical CP violation in $N_1$ decays [Eq.~(\ref{eq:esp2apr})] in terms of $w=M_{\mathrm{T}}^2/M_{\mathrm{N}_1}^2$ for $\alpha=\pi/6,\pi/4,\pi/3$, $x=M^2_{\mathrm{N}_2}/M^2_{\mathrm{N}_1}=10$, and $O(y)=10^{-3}$.}
\label{figeps2}
\end{figure}

Let us consider the very simplified limit, when $y^\nu$ and $y^t$ have all real entries and CP violation comes only from $y^{_\Sigma}$. In this case all CP violation vanishes except $\varepsilon^{(2)}_1$ in Eq.~(\ref{eps_2}). Using Eq.~(\ref{eq:case2}) with a null imaginary part for $y^\nu$ 
and considering $y^\nu\sim y^t\sim(m_\textrm{hy}/v_{_\Sigma})\sim\mathcal{O}(y)$ and $\alpha=\alpha_1=\alpha_2$, we can get the CP in $N_1$ decays as
\begin{equation}\label{eq:esp2apr}
\varepsilon _{1}^{\left( 2\right) }\simeq \frac{3}{4\pi }\frac{\sin 2\alpha
}{\left( 2+w+2\sqrt{w}\cos 2\alpha \right) }\frac{\sqrt{w}\left( 1-w\right)
}{\left( 1+\sqrt{x}\right) }\mathcal{O}(y^{2}).
\end{equation}
The corresponding numerical result of Eq.~(\ref{eq:esp2apr}) for $x=M^2_{\mathrm{N}_2}/M^2_{_{\textrm{N1}}}=10$ is shown in Fig.~\ref{figeps2} with a close-up where a maximum is obtained.

For leptogenesis to be successful it is important that the model reproduces the observed baryon-to-photon ratio~\cite{wmap},
\begin{equation}\label{etaWmap}
 \eta_B=(6.19\pm0.14)\times 10^{-10}
\end{equation}
This can be analyzed by the out of equilibrium of Majorana decays. The out of equilibrium in $N_1$ decays is controlled by the decay parameter,
\begin{equation}\label{eq:k1}
K_{\mathrm{N}_1}=\frac{\Gamma _{\mathrm{N}_1}}{H\left( T=M_{\mathrm{N}_1}\right) },
\end{equation}%
where $H\left( T\right) =1.66g_{\ast }^{1/2}T^{2}/m_{\textrm{Pl}}$ is the Hubble
parameter, $m_{\textrm{Pl}}=1.22\times 10^{19}$GeV is the Planck mass, and $g_{\ast
}\simeq 110$ in the present model is the number of relativistic degrees of freedom in the thermal bath. The total decay width is given by Eq.~(\ref{GamTot}). An estimate of $\eta_B$ can be obtained when the photon production is the standard until the recombination era as follows~\cite{Buchmuller:2004nz}
\begin{equation}\label{eq:etaC}
\eta _{B}\simeq 10^{-2}\kappa_{1}\varepsilon _{1}^{\left( 2\right) },
\end{equation}
where the final expression for the efficiency factor, $\kappa_{1}$, has a very simplified form in the strong washout regime ($K\gg1$),
\begin{equation}\label{eq:k1a}
\kappa_1\left( K\right) \simeq \frac{2}{z_{B}K}\left( 1-e^{-\frac{1}{2}z_{B}\left(
K\right) K}\right),
\end{equation}
with the parameter $z_B$,
\begin{equation}\label{eq:zb}
z_{B}\left( K\right) \simeq 2+4K^{0.13}e^{-2.5/K}.
\end{equation}%

\begin{figure}[th]
\centering
\includegraphics[height=0.6\columnwidth]{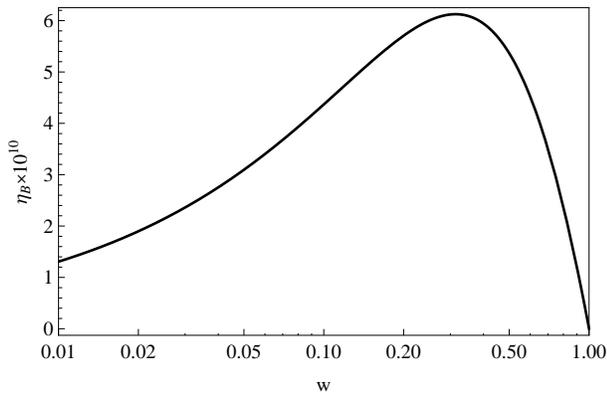}
\caption{Numerical result for the baryon-to-photon ratio $\eta_B$ as a function of $w=M_{\mathrm{T}}^2/M_{\mathrm{N}_1}^2$, obtained from Eq.~(\ref{eq:etaC}) and Eq.~(\ref{eq:esp2apr}) for $M_{\mathrm{N}_1}=3.5\times 10^{13}GeV$, $\alpha=\pi/3$, $x=M^2_{\mathrm{N}_2}/M^2_{\mathrm{N}_1}=10$, and $O(y)=0.5$.}
\label{eta1}
\end{figure}

Equation~(\ref{eq:k1a}) is calculated independently of the initial abundance of the heavy neutral fermions and, using Eq.~(\ref{eq:zb}), it may be approximated by $\kappa_{1}\left( K_{1}\right) \simeq
0.5/K_{1}^{1.2}$~\cite{DiBari:2015oca}. Without assumptions of any low-energy constraints on Yukawa couplings, let us roughly assume $x=10$ and $M_{\mathrm{N}_1}\simeq3.5\times10^{13}$GeV with the $CP$ given by Eq.(\ref{eq:esp2apr}) and maximized for $\alpha=\pi/3$. Using $O(y)\sim0.5$, we get $\eta_B^{\mathrm{max}}\simeq 6.1 \times 10^{-10}$ for $w\simeq0.3$ ($M_{\mathrm{T}}\simeq1.9\times10^{13}$GeV). It agrees with the experimental value in Eq.~(\ref{etaWmap}). Our numerical estimate for $\eta_B$ using Eq.~(\ref{eq:etaC}) is shown in Fig.~\ref{eta1} as a function of $w$.
The asymmetry generated in $N_1$ decays can be erased by the late decay of $T$, and a
further comparative analysis on a full set of Boltzmann equations to study the efficiency factors could elucidate this question about the final baryon asymmetry~\cite{DiBari:2005st}.

\section{Conclusion}
\label{sec:conclusion}

We have considered a new source of CP violation in the interplay between the type-I and type-III seesaw mechanisms. Two minimal models with one fermion triplet plus one and two fermion singlets were studied. We have shown that the CP asymmetry generated can be enhanced to produce the expected matter-antimatter asymmetry through the leptogenesis mechanism. The main feature is due to the new Majorana phase $\alpha$ which is responsible for the new CP asymmetry and cannot be absorbed by field redefinitions in the Lagrangian, as well as the fact that the heavy neutrinos cannot be in a natural diagonal basis after $\Sigma$ gets a VEV. Other connections of $\alpha$ with phenomenological constraints may or may not discard the present model. Hence, a low-energy connection should be necessary to restrict the model. Yet,  the acceptable range of $\sin \theta_{13}$ should be used to restrict the $\varepsilon^{(1)}_1$ and also to better understand the $\varepsilon^{(1)}_2$, which also may depend on $\sin \theta_{13}$ if the imaginary part of standard Yukawa couplings is nonvanishing. There is no straightforward relation to low-energy parameters since the Lagrangian in Eq.~(\ref{eq:LagMin}) (especially with more than two N's and/or T's) mixes all heavy and light eigenstates, making necessary the reconstruction of the leptonic matrix mixing in terms of known neutrino data.

By a simplified analysis we have calculated the relevant CP asymmetry considering only unflavored leptogenesis and we observed that a sufficient amount of CP violation can be obtained even if all Yukawa couplings are real, whereas $\alpha_i \ne 0$. A sufficient baryon-to-photon ratio is obtained if we consider a heavy $N_1$ ($M_{\mathrm{N}_1}\gtrsim 10^{13}$GeV). This bound can be lowered if one considers the different initial abundances for $N_1$ and their connection with low-energy parameters by Yukawa couplings~\cite{Strumia:2006qk}.
It is required to avoid the so-called cosmological gravitino problem, which occurs if our scheme is embedded in a supergravity inflation theory, where the reheating temperature has an upper bound $T_{\mathrm{reh}}\lesssim 10^{6}-10^{9}$~\cite{Khlopov:1984pf}.
The flavored investigation should point out some new aspects of leptogenesis in the context of the type I+III seesaw mechanism. In this way, the study of leptogenesis in second heavier neutral fermion ($N_2$-dominated) scenario may appropriately constrain this model.

Finally, let us remark that the present study was made with only one fermion triplet. The inclusion of a second fermion triplet will give rise to new graphs and interferences which could be very tricky. The new graphs contributing to CP asymmetry arise due to the propagator of fermion triplet components and final states as well as scalar triplets. Yet, a better understanding of this mechanism would be given when the broken phase is considered and $\mathbf{M}$ is not diagonal in general. This modifies the mechanism presented as we rewrite the Lagrangian in the mass eigenstates and may increase the final CP asymmetry.

\acknowledgments The author thanks Ricardo Gonz\'alez Felipe for enlightening discussions on the subject and for the critical reading of the manuscript. He also thanks the CFTP (IST, Lisbon) for its kind hospitality during the beginning of this work. This work was partially supported by CNPq through the Grant No. 150416/2011-3 and by the EU project MRTN-CT-2006-035505.



\bigskip

\begin{thebibliography}{99}
\bibitem{seesaw:I} P.~Minkowski,
Phys.\ Lett.\ B \textbf{67}, 421 (1977); 
T. Yanagida, in \emph{Proceedings of the Workshop on Unified Theory and
Baryon Number in the Universe}, edited by O. Sawada and A. Sugamoto (KEK,
Tsukuba, Japan, 1979); S. L. Glashow, in \emph{Quarks and Leptons, Carg\`{e}%
se 1979}, edited by M. L\'{e}vy \emph{et al}. (Plenum, New York, 1980), p. 707; M.
Gell-Mann, P. Ramond and R. Slansky, in \emph{Supergravity}, edited by . D.
Freedman and P. van Nieuwenhuizen (North Holland, Amsterdam, 1979), p. 315;
R.~N.~Mohapatra and G.~Senjanovic,
Phys.\ Rev.\ Lett.\ \textbf{44}, 912 (1980). 


\bibitem{Sakharov:1967dj}
  A.~D.~Sakharov,
  Pis'ma Zh.\ Eksp.\ Teor.\ Fiz.\  {\bf 5}, 32 (1967)
  [JETP Lett.\  {\bf 5}, 24 (1967)]
  [Sov.\ Phys.\ Usp.\  {\bf 34}, 392 (1991)]
  [Usp.\ Fiz.\ Nauk {\bf 161}, 61 (1991)].

\bibitem{sphaleron} 
G.~'t Hooft,
Phys.\ Rev.\ D \textbf{14}, 3432 (1976); Phys.\ Rev.\ D \textbf{18}, 2199 (E)
(1978)]; 
G.~'t Hooft, 
Phys.\ Rev.\ Lett.\ \textbf{37}, 8 (1976); 
F.~R.~Klinkhamer and N.~S.~Manton,
Phys.\ Rev.\ D \textbf{30}, 2212 (1984); 
V.~A.~Kuzmin, V.~A.~Rubakov and M.~E.~Shaposhnikov,
Phys.\ Lett.\ B \textbf{155}, 36 (1985). 


\bibitem{Fukugita:1986hr} M.~Fukugita and T.~Yanagida,
Phys.\ Lett.\ B \textbf{174}, 45 (1986). 

\bibitem{Davidson:2008bu} For a recent review on a leptogenesis mechanism, see
  S.~Davidson, E.~Nardi and Y.~Nir,
  Phys.\ Rept.\  {\bf 466}, 105 (2008)
  [arXiv:0802.2962 [hep-ph]].


\bibitem{Branco:2011zb} For a recent review on CP violation in the lepton sector,
see G.~C.~Branco, R.~G.~Felipe and F.~R.~Joaquim,
Rev.\ Mod.\ Phys.\ \textbf{84}, 515 (2012) [arXiv:1111.5332 [hep-ph]].



\bibitem{seesaw:II} 
M.~Magg and C.~Wetterich, 
Phys.\ Lett.\ B \textbf{94}, 61 (1980); 
J.~Schechter and J.~W.~F.~Valle,
Phys.\ Rev.\ D \textbf{22}, 2227 (1980); 
R.~N.~Mohapatra and G.~Senjanovi\'{c},
Phys.\ Rev.\ D \textbf{23}, 165 (1981). 

\bibitem{Foot:1988aq} R.~Foot, H.~Lew, X.~G.~He and G.~C.~Joshi,
Z.\ Phys.\ C \textbf{44}, 441 (1989). 



\bibitem{Branco:2002xf}
  G.~C.~Branco, R.~Gonzalez Felipe, F.~R.~Joaquim, I.~Masina, M.~N.~Rebelo and C.~A.~Savoy,
  Phys.\ Rev.\ D {\bf 67}, 073025 (2003)
  [hep-ph/0211001];
  P.~H.~Gu, H.~Zhang and S.~Zhou,
  Phys.\ Rev.\ D {\bf 74}, 076002 (2006)
  [hep-ph/0606302].

\bibitem{Frampton:2002qc}
  P.~H.~Frampton, S.~L.~Glashow and T.~Yanagida,
  Phys.\ Lett.\ B {\bf 548}, 119 (2002)
  [hep-ph/0208157];
  M.~Raidal and A.~Strumia,
  Phys.\ Lett.\ B {\bf 553}, 72 (2003)
  [hep-ph/0210021];
  V.~Barger, D.~A.~Dicus, H.~J.~He and T.~j.~Li,
  Phys.\ Lett.\ B {\bf 583}, 173 (2004)
  [hep-ph/0310278];
  R.~Gonzalez Felipe, F.~R.~Joaquim and B.~M.~Nobre,
  Phys.\ Rev.\ D {\bf 70}, 085009 (2004)
  [hep-ph/0311029];
  A.~Ibarra and G.~G.~Ross,
  Phys.\ Lett.\ B {\bf 591}, 285 (2004)
  [hep-ph/0312138];
  W.~l.~Guo, Z.~z.~Xing and S.~Zhou,
  Int.\ J.\ Mod.\ Phys.\ E {\bf 16}, 1 (2007)
  [hep-ph/0612033].


\bibitem{Branco:2001pq} G.~C.~Branco, T.~Morozumi, B.~M.~Nobre and
M.~N.~Rebelo,
Nucl.\ Phys.\ B \textbf{617}, 475 (2001) [hep-ph/0107164].

\bibitem{Bajc:2006ia}
  B.~Bajc and G.~Senjanovic,
  JHEP {\bf 0708}, 014 (2007)
  [hep-ph/0612029];
  B.~Bajc, M.~Nemevsek and G.~Senjanovic,
  Phys.\ Rev.\ D {\bf 76}, 055011 (2007)
  doi:10.1103/PhysRevD.76.055011
  [hep-ph/0703080].

    \bibitem{Kannike:2011fx}
  K.~Kannike and D.~V.~Zhuridov,
  JHEP {\bf 1107}, 102 (2011)
  [arXiv:1105.4546 [hep-ph]].

  \bibitem{su5}
  I.~Dorsner and P.~Fileviez Perez,
  JHEP {\bf 0706}, 029 (2007)
  [hep-ph/0612216];
  P.~Fileviez Perez,
  Phys.\ Lett.\ B {\bf 654}, 189 (2007)
  [hep-ph/0702287];
  P.~Fileviez Perez,
  Phys.\ Rev.\ D {\bf 76}, 071701 (2007)
  [arXiv:0705.3589 [hep-ph]];
  D.~Emmanuel-Costa, E.~T.~Franco and R.~Gonzalez Felipe,
  JHEP {\bf 1108}, 017 (2011)
  [arXiv:1104.2046 [hep-ph]];


\bibitem{Sierra:2013ypa}
  D.~Aristizabal Sierra, I.~de Medeiros Varzielas and E.~Houet,
  Phys.\ Rev.\ D {\bf 87}, no. 9, 093009 (2013)
  [arXiv:1302.6499 [hep-ph]];
  D.~Aristizabal Sierra and I.~de Medeiros Varzielas,
  JHEP {\bf 1407}, 042 (2014)
  [arXiv:1404.2529 [hep-ph]].

\bibitem{DiBari:2005st}
  P.~Di Bari,
  Nucl.\ Phys.\ B {\bf 727}, 318 (2005)
  [hep-ph/0502082];
  O.~Vives,
  Phys.\ Rev.\ D {\bf 73}, 073006 (2006)
  [hep-ph/0512160];
  G.~Engelhard, Y.~Grossman, E.~Nardi and Y.~Nir,
  Phys.\ Rev.\ Lett.\  {\bf 99}, 081802 (2007)
  [hep-ph/0612187].

\bibitem{Mohapatra:1986bd}
  R.~N.~Mohapatra and J.~W.~F.~Valle,
  Phys.\ Rev.\ D {\bf 34}, 1642 (1986);
  M.~C.~Gonzalez-Garcia and J.~W.~F.~Valle,
  Phys.\ Lett.\ B {\bf 216}, 360 (1989);
  F.~Deppisch and J.~W.~F.~Valle,
  Phys.\ Rev.\ D {\bf 72}, 036001 (2005)
  [hep-ph/0406040].

\bibitem{Branco:2006ce}
  G.~C.~Branco, R.~Gonzalez Felipe and F.~R.~Joaquim,
  Phys.\ Lett.\ B {\bf 645}, 432 (2007)
  [hep-ph/0609297];
  S.~Pascoli, S.~T.~Petcov and A.~Riotto,
  Phys.\ Rev.\ D {\bf 75}, 083511 (2007)
  [hep-ph/0609125].
  S.~Pascoli, S.~T.~Petcov and A.~Riotto,
  Nucl.\ Phys.\ B {\bf 774}, 1 (2007)
  [hep-ph/0611338].
  E.~Molinaro and S.~T.~Petcov,
  Phys.\ Lett.\ B {\bf 671}, 60 (2009)
  [arXiv:0808.3534 [hep-ph]].

\bibitem{Covi:1996wh} L.~Covi, E.~Roulet and F.~Vissani,
Phys.\ Lett.\ B \textbf{384}, 169 (1996) [hep-ph/9605319].

\bibitem{Buchmuller:2004nz} W.~Buchmuller, P.~Di Bari and M.~Plumacher,
Annals Phys.\ \textbf{315}, 305 (2005) [hep-ph/0401240].



\bibitem{flavored}
  R.~Barbieri, P.~Creminelli, A.~Strumia and N.~Tetradis,
  Nucl.\ Phys.\ B {\bf 575}, 61 (2000)
  [hep-ph/9911315];
  T.~Endoh, T.~Morozumi and Z.~h.~Xiong,
  Prog.\ Theor.\ Phys.\  {\bf 111}, 123 (2004)
  [hep-ph/0308276];
  A.~Pilaftsis and T.~E.~J.~Underwood,
  Phys.\ Rev.\ D {\bf 72}, 113001 (2005)
  [hep-ph/0506107];
  A.~Abada, S.~Davidson, F.~X.~Josse-Michaux, M.~Losada and A.~Riotto,
  JCAP {\bf 0604}, 004 (2006)
  [hep-ph/0601083];
  E.~Nardi, Y.~Nir, E.~Roulet and J.~Racker,
  JHEP {\bf 0601}, 164 (2006)
  [hep-ph/0601084];
  A.~Abada, S.~Davidson, A.~Ibarra, F.-X.~Josse-Michaux, M.~Losada and A.~Riotto,
  JHEP {\bf 0609}, 010 (2006)
  [hep-ph/0605281];
  P.~S.~Bhupal Dev, P.~Millington, A.~Pilaftsis and D.~Teresi,
  Nucl.\ Phys.\ B {\bf 886}, 569 (2014)
  [arXiv:1404.1003 [hep-ph]].

\bibitem{wmap}
  C.~L.~Bennett {\it et al.} [WMAP Collaboration],
  Astrophys.\ J.\ Suppl.\  {\bf 208}, 20 (2013)
  [arXiv:1212.5225 [astro-ph.CO]].

%

\bibitem{DiBari:2015oca}
  P.~Di Bari and S.~F.~King,
  arXiv:1507.06431 [hep-ph].

\bibitem{Strumia:2006qk}
  A.~Strumia,
  Lectures given at the Les Houches Summer School on Theoretical Physics: Session
84: Particle Physics Beyond the Standard Model,
  hep-ph/0608347.

\bibitem{Khlopov:1984pf}
  M.~Y.~Khlopov and A.~D.~Linde,
  Phys.\ Lett.\ B {\bf 138}, 265 (1984);
  J.~R.~Ellis, J.~E.~Kim and D.~V.~Nanopoulos,
  Phys.\ Lett.\ B {\bf 145}, 181 (1984);
  M.~Kawasaki, K.~Kohri, T.~Moroi and A.~Yotsuyanagi,
  Phys.\ Rev.\ D {\bf 78}, 065011 (2008)
  [arXiv:0804.3745 [hep-ph]].







\end{thebibliography}
\end{document}